\documentclass[prl,twocolumn]{revtex4-1}

\usepackage{graphicx}   
\usepackage{hyperref}   

\begin{document}

\title{Spatiotemporally complete condensation in a non-Poissonian exclusion process}
\author{Robert~J.~Concannon}
\author{Richard~A.~Blythe}
\affiliation{SUPA, School of Physics and Astronomy, University of Edinburgh, Mayfield Road, Edinburgh EH9 3JZ, UK}
\date{10th January 2014}

\begin{abstract}
We investigate a non-Poissonian version of the asymmetric simple exclusion process, motivated by the observation that  coarse-graining the interactions between particles in complex systems generically leads to a stochastic process with a non-Markovian (history-dependent) character.  We characterize a large family of one-dimensional hopping processes using a waiting-time distribution for individual particle hops.  We find that when its variance is infinite, a real-space condensate forms that is complete in space (involves all particles) and time (exists at almost any given instant) in the thermodynamic limit.  The mechanism for the onset and stability of the condensate are both rather subtle, and depends on the microscopic dynamics subsequent to a failed particle hop attempts. 
\end{abstract}

\maketitle


When modeling complex interacting systems that are out of equilibrium, it is customary within the statistical physics community \cite{Schmittmann95,Privman97,Schutz01,Helbing01,Henkel08,Schadschneider11} to couch their dynamics in terms of Poisson processes. That is, given the current state of the system, a transition to a new state is assumed to occur at a constant rate. There are, however, situations where this does not accurately reflect reality. For example, in stochastic models of infectious diseases \cite{Andersson00}, the recovery rate of an individual may peak at some characteristic infection time \cite{Black09,Brett13}. Likewise, each step of a motor protein can be viewed as the consequence of a sequence of internal processes (see e.g.~\cite{Cross04,Nishinari05,Basu07}), which leads to a non-Poissonian hopping dynamics. Indeed, whenever a stochastic dynamics is obtained by coarse-graining a more complete description, Poisson processes will arise as an exception rather than the rule.

The most basic interaction between stochastic processes is to prevent events from occurring: e.g., an infected agent inhibiting the recovery of another, or one motor protein blocking another from moving. A fundamental question then arises: How do the microscopic consequences of a blocked transition affect the macroscopic properties of the system? For Poisson processes the answer is clear: whenever a transition can occur, it occurs at a constant rate. By contrast, the answer for non-Poissonian processes is unclear due to limited analytical techniques for analysing interacting non-Markovian systems. Consequently these have rarely been studied by statistical physicists (Ref.~\cite{Hirschberg09} is a notable exception).

Here we make progress towards an answer in the context of the asymmetric simple exclusion process (ASEP) on a ring \cite{Derrida98,Blythe07}, in which hard-core particles hop in a preferred direction on a (periodic) one-dimensional lattice. The ASEP and its variants underpin many applications of interest, e.g., biopolymerisation \cite{asepbiopolymers}, traffic flow \cite{Helbing01} and molecular motors \cite{Schadschneider11}. When formulated as Poisson processes, their stationary states are well characterized, which has led to a deep understanding such nonequilibrium effects as phase transitions and spontaneous symmetry breaking \cite{Blythe07}. Most pertinently, the stationary state of the Poissonian ASEP on a ring is wholly unremarkable: all microscopic configurations are equally likely and the macrostate is a fluid (see e.g.~\cite{Derrida98,Schadschneider11}).

Here, we report that in a non-Poissonian ASEP where the distribution of times between particle hop attempts and has an infinite variance, the stationary state changes dramatically to reveal a distinctive condensation phenomenon. In this state, all particles coalesce into a single solid block at almost all times: the condensation is \emph{complete} not just in space \cite{Jeon00,Evans06} but also in time.  This condensate forms through an ageing process whereby successive particles are immobile for increasing lengths of time. This differs from the more usual real-space condensation seen in zero-range processes which arises through a current balance between the condensate and the background fluid \cite{Evans05}. It is also distinct from the condensate previously seen in the ASEP with disordered hop-rates \cite{Krug96,Evans96} in that it is immobile at almost all times. As we discuss below, our main analytical tool for understanding the subtle properties of the condensate is to integrate out failed hop attempts. This approach delivers an accelerated simulation method for interacting non-Poissonian systems which may generalize to related model systems.

\begin{figure}
\centerline{\includegraphics[width=0.9\linewidth]{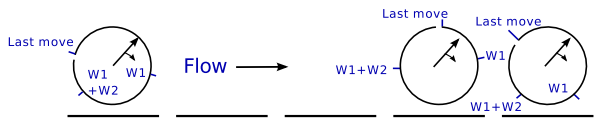}}
\caption{\label{fig:asepcartoon}(Color online) Hop times in the non-Poissonian ASEP. Each particle is represented as a clock showing the current time (arrow), the last time each particle moved (break in the circle) and the times of the next hop attempts (short lines). These are given by the cumulative waiting times $W1, W1+W2, \ldots$, where each term is a random variable drawn from $p(W)$. In this figure, the first hop attempts of the outer particles succeed, the first attempt of the middle particle fails but its second attempt succeeds.}
\end{figure}

We begin by defining the non-Poissonian ASEP. It comprises $N$ particles on a one-dimensional lattice of $L$ sites with periodic boundary conditions, with at most one particle on any given site.  At time $t=0$, each particle $i$ is independently assigned a waiting time $W_i$ from the distribution $p(W_i)$.  When a particle's waiting time expires, it attempts to hop to the site to its right, succeeding only if the receiving site is empty.  Whether or not the  attempt is a success, the particle is assigned a new waiting time from $p(W_i)$. Then, the next hop attempt  takes place at time $t+W_i$, where $t$ is the time of the most recent hop attempt. Fig.~\ref{fig:asepcartoon} shows a schedule of hop attempts.

This formulation amounts to restarting the process that generates the waiting-time distribution $p(W)$ when a hop attempt fails. This scheme was used in \cite{Gorissen12} to model motor proteins. In that work, the waiting-time distribution had an exponential tail: under such conditions we find that the non-Poissonian ASEP on a ring has broadly the same behaviour as the Poissonian version.

It is also possible for a waiting-time distribution to have a non-exponential tail. For example, in Bouchaud's trap model  \cite{Bouchaud92,Monthus96} particles fall into traps with exponentially-distributed depths $\epsilon$. Particles escape from their traps as a Poisson process with a rate given by the Arrhenius law ${\rm e}^{-\epsilon/T}$ where $T$ is the temperature. The waiting-time distribution then has a power-law tail $p(W)\sim W^{-(T+1)}$. Such models have been used to describe ageing phenomena in glasses \cite{Monthus96} and kink propagation along a dislocation line \cite{Feigelman88}. The non-Poissonian ASEP with a power-law waiting-time distribution can therefore be viewed as a trap model where particles continually hop between traps in some energy landscape, but move in a physical space only when the site in front is unoccupied at the instant a new trap is entered.

\begin{figure}
\centerline{\includegraphics[width=\linewidth]{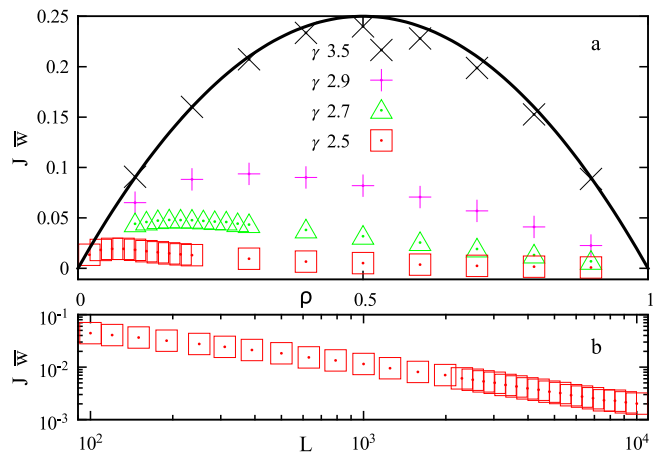}}
\caption{\label{fig:funddiag}(Color online) Flux, $J\bar{W}$, as a function of density, $\rho$, at fixed system size $L=500$ (upper figure) and as a function of system size with $\rho=0.1$, $\gamma=2.5$ (lower figure) as determined from direct simulations of the hopping dynamics.}
\end{figure}

An initial study of the ASEP with a power-law waiting-time distribution, $p(W)=(\gamma-1)W^{-\gamma}$ for $\gamma>1$, $W>1$, is facilitated by a standard Monte Carlo waiting-time algorithm \cite{Gillespie77}. The fundamental diagram---a plot of the steady-state flux, $J\bar{W}$, against particle density, $\rho=N/L$---is shown in the upper part of Fig.~\ref{fig:funddiag} for various $\gamma$ where $J$ is the particle current and $\bar{W}$ is the mean of $p(W)$.  For Poissonian dynamics, it is known that $J \bar{W} = \rho(1-\rho)$ \cite{Derrida98,Schadschneider11}.  For $\gamma>3$, we see reasonable agreement with this result.  For $\gamma<3$, significant deviations are apparent.  First, the flux is no longer symmetric about $\rho=\frac{1}{2}$: this is due to breaking of the particle-hole symmetry that is present in the Poissonian case. Secondly, the current is somewhat smaller, indeed apparently vanishing in the thermodynamic limit ($L\to\infty, N\to\infty$ with $\rho = N/L$ fixed) as the lower part of Fig.~\ref{fig:funddiag} shows.

\begin{figure}
\centerline{\includegraphics[width=\linewidth]{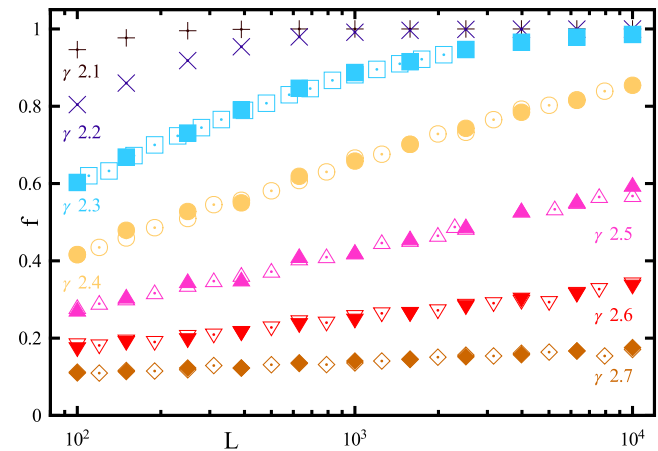}}
\caption{\label{fig:fplot}(Color online) Fraction, $f$, of time  during which a single solid block is present obtained from simulations using the direct algorithm (open symbols for $\gamma\ge2.3$) and the accelerated algorithm (solid symbols for $\gamma\ge2.3$ and all data for $\gamma<2.3$). In all cases the density $\rho=0.1$.}
\end{figure}

When $\gamma<3$, the variance of the waiting-time distribution is infinite. Intuition then suggests that anomalously large waiting times are generated sufficiently often that \emph{all} the particles can become blocked by a single, immobile particle at the head of the block. If these solid blocks are present for a fraction of time, $f$, that increases with system size, $L$, the current would vanish as $L\to\infty$. The function $f(L)$, as determined by direct Monte Carlo simulations, is shown in Fig.~\ref{fig:fplot} as open symbols.  The data show some increase of $f$ with $L$ but, over the range of system sizes and $\gamma$ that are accessible with this method, it is not clear whether $f$ will approach unity in the thermodynamic limit or saturate at some smaller value.

\begin{figure}
\centerline{\includegraphics[width=\linewidth]{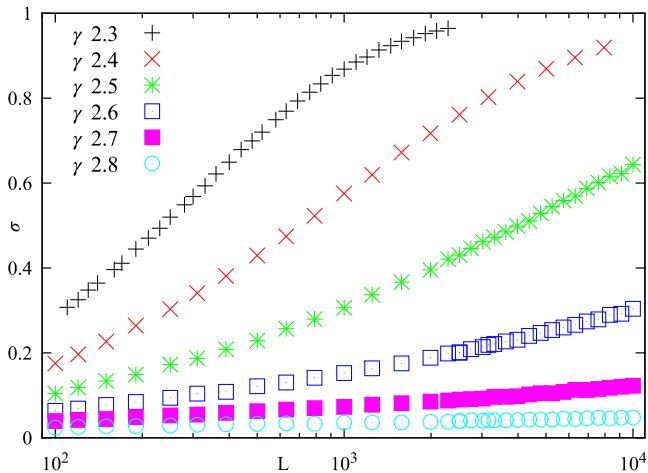}}
\caption{\label{fig:sigplot}(Color online) Fraction, $\sigma$, of condensates that do not fully dissolve as a function of system size. All data obtained using the direct simulation algorithm.}
\end{figure}

We now argue that across the entire range $2<\gamma<3$, the fraction of time that a condensate is present approaches unity in the thermodynamic limit, and identify the physical mechanism behind this effect.  The first step is to recognize that once a single solid block is formed, it need not fully dissolve before the next one is formed: some particles may break off from the front of the block, traverse the system, and rejoin the back of the block while the rest of the block remains stationary. Fig.~\ref{fig:sigplot} shows that the fraction, $\sigma$, of condensates that are formed out of a remnant of the last increases steadily with system size. The lifetime of a current-carrying state (the \emph{fluid}) can be estimated as that required for a particle to traverse the system, i.e., a time of order $L$. The  lifetime of the condensate, meanwhile, requires a more detailed analysis.

The crucial physical component of this analysis is elucidated by revisiting the trap picture. Here it is known \cite{Monthus96} that, in the stationary state, the waiting time $W$ that was most recently assigned to any randomly-chosen particle has a distribution with a ${\sim}W^{1-\gamma}$ tail, as opposed to the original $p(W)\sim W^{-\gamma}$ distribution.  Consider now a pair of particles on neighboring sites where the first particle is about to move off and the second arrived a time $T$ ago. As we will show below, the \emph{residual waiting time}, $\Delta$, at which the second particle moves off relative to the first has a hybrid of these two distributions, crossing over from the slower $\Delta^{1-\gamma}$ decay to the $\Delta^{-\gamma}$ decay at $\Delta \sim T$.  This is because, from the perspective of the first particle, the second is \emph{not} a randomly-chosen particle. Rather, their motions are correlated because the second particle must have attempted to hop (whether successful or not) since the first particle last hopped. As we will see, this correlation leads to $\Delta$ growing with $T$ in such a way that the condensate persists for a time that grows faster than the $O(L)$ fluid lifetime in the thermodynamic limit.


We thus define $p(\Delta|T)$ as the residual waiting time distribution for a particle that is unable to move for a time $T$ after arriving at a site, due to the presence of a particle in front.  It can be written as $p(\Delta|T)=\int_{0}^{T}{\rm d} t \mu(t) p(T+\Delta-t)$ where $\mu(t)$ is the density of hop attempts at time $t$ conditioned on an attempt taking place at time $t=0$.  It seems reasonable that $\mu(t)$ will approach the constant $1/\bar{W}$ as $t\to\infty$, where we recall that $\bar{W}$ is the mean time between attempts.  One can show  this intuition holds by appealing to the Laplace transform $\tilde{\mu}(s) = [1 - \tilde{p}_1(s)]^{-1}$  where $\tilde{x}(s)$ denotes the Laplace transform of a function $x(t)$ \cite{Feller2}. The dominant singularity in $\tilde{\mu}(s)$ is a pole at the origin with amplitude $1/\bar{W}$, yielding the expected $t\to\infty$ behavior. Approximating $\mu(t)$ by this constant value for \emph{all} $t$ in the integral for $p(\Delta|T)$, we find that
\begin{equation}
\label{PWT}
p(\Delta|T) \sim \frac{\gamma-2}{1 - T^{2-\gamma}} \left[\Delta^{1-\gamma}-(T+\Delta)^{1-\gamma} \right]  \;,
\end{equation}
where the prefactor ensures correct normalization. We expect (\ref{PWT}) to hold for $T$ sufficiently large that transients in $\mu(t)$ can be neglected: an analysis of the Laplace transform of $p(\Delta|T)$ shows that this is indeed the case \cite{RobPhD}.


The result (\ref{PWT}) allows us to accelerate the Monte Carlo simulations, thereby gaining access to larger systems and smaller values of $\gamma$ than were possible with the direct approach.  This is achieved by noting that when a particle arrives at a site where it cannot move, the time $T$ for which it must wait before the particle in front moves off is known. When $T$ is large enough for the asymptotic result (\ref{PWT}) for $p(W|T)$ to hold, it can be used to assign the arriving particle's residual waiting time in a single operation instead of repeatedly resampling the original waiting time distribution until the sum of all such times exceeds $T$.  For smaller $T$, we fall back on the latter direct-summation approach to retain accuracy. Empirically, we have found that across the whole range of $\gamma$, the distribution (\ref{PWT}) can be used for $T> 2000$. In Fig.~\ref{fig:fplot}, data from this accelerated algorithm (filled symbols) agree well with those from the brute-force algorithm (open symbols).  We further see for $\gamma\le2.3$ that the fraction of time spent in a complete condensate, $f(L)$, reaches unity as $L$ increases. For larger $\gamma$, $f(L)$ steadily increases, suggesting that the same asymptote will be seen here too.


The result (\ref{PWT}) also allows analytical progress. First, we can estimate the probability, $\sigma$, that a condensate does not completely dissolve before the next is formed.  For dissolution to occur, all particles must receive a residual waiting time $\Delta_i$ smaller than the time, $C_{i-1}$, taken by the particle in front to rejoin the back of the condensate.  Assuming that the times $C_{i}$ are of order $L$ (since this is the number of sites that must be traversed), and that all blocking times $T_i$ are all of order $L$ (since this is what is required for the condensate to form in the first place), we have $1-\sigma = \prod_{i} [ 1 - \int_{C_{i-1}}^{\infty} {\rm d}\Delta_i P(\Delta_i|T_i) ] \sim {\rm e}^{-aL^{3-\gamma}}$ for some $a>0$. When $\gamma<3$, the fraction $\sigma \to 1$ as $L\to\infty$, showing that typically each condensate comprises a remnant of the previous condensate. This is consistent with the empirical data for $\sigma$ shown in Fig.~\ref{fig:sigplot}.

We now consider a particle that has made the round trip from the front to the back of the condensate. Let us index the particles with $i=1,2,\ldots$ from the front to the back of the condensate.  Relative to the time at which particle $N$ left the previous condensate, the time that particle $N-1$ will move is given by the sum of the residual waiting times $\Delta_j$ for particles $j=1,2,\ldots,N-1$. Then, the blocking time for particle $N$ is
\begin{equation}
\label{rec}
T_N = \sum_{j=1}^{N-1} \Delta_j - C_N
\end{equation}
where $C_N$ is the time particle $N$ took to make the round-trip from the front to the back of the condensate.

After sufficiently many round trips, we assume the distributions of the random variables $T$, $\Delta$ and $C$ in (\ref{rec}) become stationary. Using (\ref{PWT}), one can show that the mean of $\Delta$, for some known value of $T$, is $\frac{T^{3-\gamma}}{(3-\gamma)(\gamma-1)} - \frac{\gamma-2}{3(2-\gamma)}$. By averaging (\ref{rec}) and making the mean-field type approximation $\overline{g(T)} = g(\bar{T})$, for any function $g(u)$, we find for large $L$ the stationary mean blocking time to be
\begin{equation}
\label{Tbar}
\bar{T} \sim \left[ \frac{\rho L}{(3-\gamma)(\gamma-1)} \right]^{\frac{1}{\gamma-2}} \;.
\end{equation}
The mean round-trip time $\bar{C}$ does not enter into this asymptotic (large-$L$) result if it grows linearly in $L$, as we have assumed throughout this work. Now let $\bar{\Delta}_{+}$ be the mean residual waiting time for a blocked particle in a stationary condensate \emph{conditioned} on it being large enough to precipitate the next condensate (i.e., $\Delta>\bar{C}$).  The lifetime of each condensate can then be estimated as $\bar{\Delta}_{+} - \bar{C}$ whose leading term scales as $L^{\chi}$ with $\chi = \frac{3-\gamma}{\gamma-2}+\gamma-2$. With the fluid lifetime being of order $\bar{C} \sim L$, we find now that there is a separation of timescales for the condensate and the fluid for $\gamma<3$, implying that a single condensate comprising all particles is present at almost all times in the thermodynamic limit.

\begin{figure}[t]
\includegraphics[width=\linewidth]{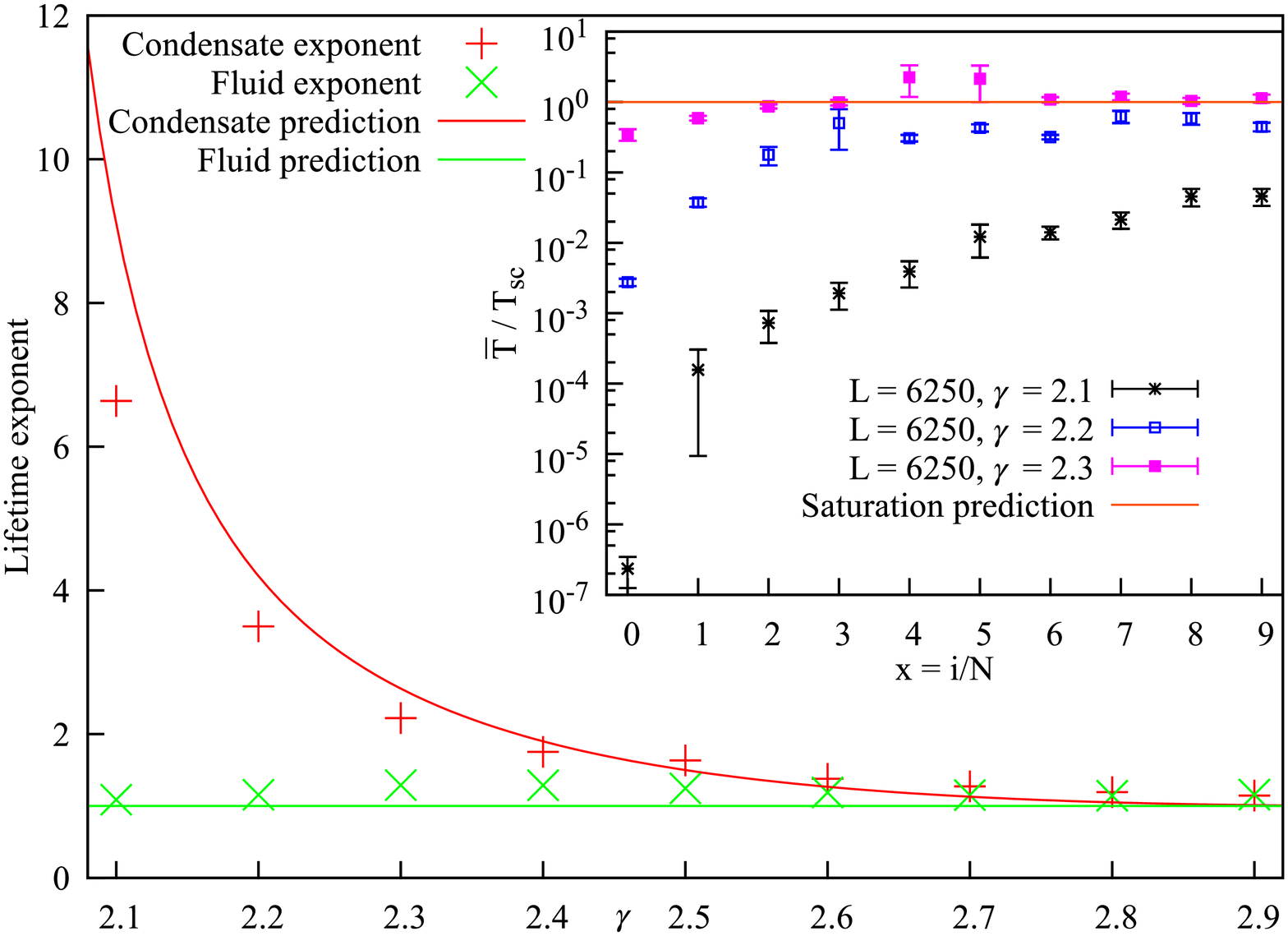}
\caption{\label{fig:exp}(Color online) Main figure: Empirical scaling exponents of the condensate and fluid lifetimes with system size for various $\gamma$ at $\rho=0.1$, along with the theoretical predictions. Inset: Approach of the blocking time $T_i$ to its stationary value, normalized to the prediction $T_{sc}$ for the latter obtained using the self-consistent sampling algorithm.}
\end{figure}

We test this picture with simulations. In the main part of Fig.~\ref{fig:exp} we show how the condensate and fluid lifetimes scale with system size.  We find that the predicted $L^\chi$ growth of the condensate lifetime is consistent with simulation. Although the fluid lifetime appears to scale slightly faster than linearly with $L$, as we have assumed in the analysis, the condensate exponent always exceeds the fluid exponent, showing that the predicted separation of timescales is indeed present.  Another test of the theory is to examine the stationary blocking time predicted by (\ref{Tbar}).  Here we find the exponent is well-predicted, but the prefactor can be several orders of magnitude larger than that observed. This discrepancy is due to assuming that the random variable $T_j$ can be replaced with its mean, $\bar{T}$.  To correct for this, we constructed the stationary distribution of $T$ self-consistently by manipulating a pool of $n=10^4$ instances of the random variable $T$. In each iteration of this algorithm, $N-1$ values are sampled with replacement from the pool, and then used in (\ref{PWT}) to generate residual waiting times $\Delta_j$. Performing the sum (\ref{rec}) yields a new instance of the random variable $T$, which replaces one of the existing values in the pool. We found that even when neglecting the round-trip time $C_i$ completely (i.e., setting it to zero), the mean blocking time $T_{sc}$ obtained after convergence is consistent with simulation (at least for large $L$, see Fig.~\ref{fig:exp} inset).

Collectively, our simulation and analytical results paint the following picture of condensation in the non-Poissonian ASEP.  At first, a current flows until one particle is assigned a waiting time large enough for the rest to catch up and form a single, solid condensate.  Each particle in the condensate is blocked for a successively longer time $T$, which in turn implies, through (\ref{PWT}), that the time $\Delta$, between a particle becoming unblocked and beginning to move, also tends to grow. This growth is sufficiently fast that one of the particles that breaks off will make the round trip and restore the condensate before the next particle can move off. This process is repeated until a stationary state is reached in which the lifetime of a single condensate grows as a faster power of system size than that of the fluid.  In this sense, the condensate is complete in both space (it involves all the particles) and time (it is present almost all the time).

We have therefore found that in the specific case of the ASEP, non-Poissonian particle hopping dynamics induce an interaction that is sufficiently strong to induce a condensation phenomenon that is completely absent in the Poissonian case. The condensation mechanism is distinct from that of related systems, such as the zero-range process. Moreover, a non-Markovian generalization of the latter \cite{Hirschberg09} shows a \emph{raising}  of the critical density for condensation, as opposed to the greater propensity towards condensation found here. Further study \cite{RobPhD} has shown the condensate to be robust to spatial asymmetry in the dynamics. On the other hand, it goes away if the process generating the power-law waiting time distribution is started \emph{only} when a particle is free to move. This demonstrates that relationship between the way in which non-Poissonian stochastic processes interact and the emergent properties of a system is not straightforward, and that further investigation of interacting non-Poissonian processes is needed to build a systematic picture of this relationship.  


\textit{Acknowledgments---} We thank Martin Evans, Stefan Grosskinsky and Bartek Waclaw for helpful comments, and Peter Sollich for pointing out the connection to the Bouchaud trap model. This work has made use of the resources provided by the Edinburgh Compute and Data Facility (ECDF) (http://www.ecdf.ed.ac.uk/). RJC thanks the EPSRC for financial support.

\end{document}